\pdfoutput=1
\documentclass[a4paper]{amsart}

\usepackage[T1]{fontenc}
\usepackage{lmodern}
\usepackage{graphicx}
\usepackage{mathtools,amsthm,amssymb}
\usepackage[sort&compress,round]{natbib}

\usepackage{pgfplots}
\usepackage{pgfplotstable}
\usepgfplotslibrary{colormaps}
\pgfplotsset{compat=1.16,samples=1000}

\theoremstyle{definition}
\newtheorem{exmp}     {Example}
\newtheorem{rem}[exmp]{Remark}
\theoremstyle{plain}
\newtheorem{thm}[exmp]{Theorem}

\newcommand{\N}{\mathbb N}
\newcommand{\R}{\mathbb R}
\newcommand{\e}{\varepsilon}
\newcommand{\E}{\operatorname{E}}

\newcommand{\dconv}{ \ensuremath{ \overset{d}{\longrightarrow}}}
\newcommand{\pconv}{\ensuremath{ \overset p {\longrightarrow}}}

\newcommand{\cp}{m}
\newcommand{\I}{\ensuremath \mathcal I}
\newcommand{\adap}{\ensuremath{ \mathrm{ad} }}
\newcommand{\equi}{\ensuremath{ \mathrm{eq} }}
\newcommand{\non}{\ensuremath{ \mathrm{non} }}

\begin{document}

\title[Adaptive Quantile Computation for Brownian Bridge]%
{Adaptive Quantile Computation for Brownian Bridge in
Change-Point Analysis}

\author[J.\ Franke]{J\"urgen Franke}
\address{Fachbereich Mathematik\\
TU Kaiserslautern\\
Postfach 3049\\
67653 Kaiserslautern\\
Germany}
\curraddr{Fraunhofer ITWM\\
Fraunhofer-Platz 1\\
67663 Kaiserslautern\\
Germany}
\email{franke@mathematik.uni-kl.de}

\author[M.\ Hefter]{Mario Hefter}
\address{Fachbereich Mathematik\\
TU Kaiserslautern\\
Postfach 3049\\
67653 Kaiserslautern\\
Germany}
\email{hefter@mathematik.uni-kl.de}

\author[A.\ Herzwurm]{Andr\'e Herzwurm}
\address{R+V Versicherung AG\\
Raiffeisenplatz 2\\
65189 Wiesbaden\\
Germany}
\email{andre.herzwurm@ruv.de}

\author[K.\ Ritter]{Klaus Ritter}
\address{Fachbereich Mathematik\\
TU Kaiserslautern\\
Postfach 3049\\
67653 Kaiserslautern\\
Germany}
\email{ritter@mathematik.uni-kl.de}

\author[S.\ Schwaar]{Stefanie Schwaar}
\address{Fraunhofer ITWM\\
Fraunhofer-Platz 1\\
67663 Kaiserslautern\\
Germany}
\email{stefanie.schwaar@itwm.fraunhofer.de}

\keywords{
change-point problem,
weighted CUSUM statistic,
weighted Brownian bridge,
sup-norm quantiles,
Monte Carlo algorithm,
adaptive discretization}

\date{November 12, 2020}

\begin{abstract}
As an example for the fast calculation of distributional parameters of
Gaussian processes, we propose a new Monte Carlo algorithm for the
computation of quantiles of the supremum norm of weighted Brownian
bridges. As it is known, the corresponding distributions arise
asymptotically for weighted CUSUM statistics for change-point detection.
The new algorithm employs an adaptive (sequential) time discretization
for the trajectories of the Brownian bridge. A simulation study shows
that the new algorithm by far outperforms the standard approach, which
employs a uniform time discretization.
\end{abstract}

\maketitle


\section{Introduction}

In statistical inference, asymptotics frequently leads to the
distribution of nonlinear functionals of Gaussian processes. E.g., the
construction of uniform asymptotic confidence bands for a regression
function based on kernel estimates requires the study of the supremum
of the absolute values of a certain Gaussian process,
cf.\ \citet*[Sec.~4.3]{book:Haerdle90}. To mention another
example, for testing the equality of mean functions in functional data
analysis, a test statistic is used which under the hypothesis converges
in distribution to an integral of the square of a Gaussian process,
cf.\ \citet*[Sec.~5.1]{book:HorvKoko12}.

In some cases like the first example above,
cf.\ \citet*{BickRos73}, the distribution of the nonlinear
functional may be derived in a form which, in particular, allows for the
calculation of quantiles for tests and confidence assessments. In many
other cases, however, distributional characteristics have to be
calculated numerically by Monte Carlo simulation. Even in rather simple
cases where the Gaussian process is just a Wiener process or a
Brownian bridge, using the standard approximation of a
continuous-time Gaussian process by a corresponding discrete-time process
on an equidistant grid may result in a severe computational load if a
decent quality of approximation is required. We shall discuss this
below in more detail for a specific case.

In this paper, we show that this problem can be overcome by using fast
adaptive approximation methods for the strong (or pathwise)
approximation of nonlinear functionals of Gaussian processes.
Adaptive algorithms employ sequential strategies to construct
the underlying discretization, which may therefore be adjusted to key
features of the individual trajectories.

For illustrating our approach, we focus on weighted CUSUM tests for
change-points, which leads to the distribution of the supremum of
a weighted reflecting Brownian bridge,
i.e., the supremum norm of a weighted Brownian bridge.
We stress that the basic idea can be transferred to other situations
where, e.g., quantiles of nonlinear functionals of Gaussian processes
have to be calculated by Monte Carlo simulation.

Change-point tests are of interest in many areas of applications,
e.g., in production  monitoring, see \citet*{Page57},
on-line-monitoring of intensive-care patients, see \citet*{FriedImhoff04},
or global warming studies, see \citet*{GallagherEtAl13}, to name
just a few. The first publications about testing for a change in data
go back to the 1950s, see, e.g., \citet*{Page57}, who has considered
testing for a change in the mean and
has used weighted cumulated sums of sample residuals,
so called weighted CUSUM statistics. The corresponding weight
function is given by
\[
w(t) = (t \cdot (1-t))^{-1/2}
\]
for $0<t<1$, so that the weighted cumulated sums are
variance stable. The distribution of those statistics is
determined asymptotically in a variant of the
Darling-Erd\"os Theorem, see Theorem \ref{t1}, which immediately
yields asymptotic quantiles.

To cope with performance problems regarding size and power,
the standard weights of CUSUM statistics have been modified which
results asymptotically in the distribution of
the supremum of a weighted reflecting Brownian bridge.
Those statistics have no size problems and better power against changes
of the mean closer to the boundaries of the observation interval.
Simulation studies using different weight functions and analyzing the
power of the corresponding CUSUM-type tests for different positions of
the change (early, middle and late) show the importance of the weight
function, see \citet*{book:CsorgoHorvath},
\citet*{incoll:KirchTadjuidje16}, and \citet*{Dis:Schwaar}.
An overview to such general
CUSUM-type tests is given in \citet{AueHorvath13}.

In the present paper we consider weight functions of the form
\[
w_{\eta,\gamma}(t) = 1_{\left]\eta,1-\eta\right[}(t)
\cdot (t \cdot (1-t))^{-\gamma}
\]
for $0 < t < 1$ with parameters $0 \leq \eta < 1/2$ and
$0 \leq \gamma \leq 1/2$, and the corresponding convergence result for
$(\eta,\gamma) \neq (0,1/2)$ is formulated in Theorem \ref{t2}.
Except for the extremal cases $\gamma = 0$ and $\gamma = 1/2$,
there is no known method for
analytically calculating quantiles or other characteristics of the limit
distributions for these or more general weight functions.
Hence, we have to use Monte Carlo
simulation where a crucial part consists in generating paths of
a Brownian bridge. This is in
particular computationally very expensive if we are interested in
calculating, e.g., extreme quantiles beyond the common levels
$0.05$ or $0.01$ with a high precision up to $10^{-3}$. Such extreme
levels of confidence are common in many applications in industry or
medicine where a high degree of reliability is required. Also, in view
of the Bonferroni inequality, low $p$-values of tests are of interest
in multiple testing situations including many hypotheses, see, e.g.,
\citet*{Hochberg88}.

In this paper, we propose an adaptive algorithm which
reduces the computation time for calculating asymptotic quantiles of
CUSUM test statistics with weight function $w_{\eta,\gamma}$ for
$\gamma \not\in \{0,1/2\}$, where the reduction turns out to be
dramatic in the more challenging situations.
Consider, for instance, the task to compute the
$0.95$-quantile with accuracy $10^{-2}$ for the parameters
$\eta=0$ and $\gamma = 0.45$. On a common up-to-date processor
the standard algorithm with an equidistant grid requires more than
two hours of computation time, while
the adaptive algorithm achieves the same goal within 12 seconds.
The reason for this is, roughly speaking, that
the adaptive algorithm allows
to sample almost exactly from the correct limit distribution
at a reasonable computational cost.

This paper is organized as follows.
In Section \ref{sec:CPT} we give a brief sketch of the change-point
application that is used as a demonstrator for our approach.
In Sections~\ref{poweradap} and \ref{s3} we consider
the strong approximation of the supremum of the
unweighted Brownian motion and of weighted reflecting
Brownian bridges, respectively.
For the former problem, \citet*{MR3605752} have constructed an adaptive
algorithm that strikingly outperforms all non-adaptive algorithms,
see Theorems \ref{t3} and \ref{t4}.
See also \citet*{Calvin1997,Calvin2001,Calvin2004} for related
convergence results.
No such result is known for weighted reflecting
Brownian bridges, but still we construct a
modification of the adaptive algorithm
for these processes in Section \ref{s4.1}. Numerical experiments reveal
again the vast superiority of the adaptive algorithm over, at least, the
standard algorithm that is based on an equidistant
grid, see Section \ref{strongnumeriocs}.
In Section \ref{s5} we study quantile computation
for the supremum of a weighted reflecting
Brownian bridge. We present a new algorithm, with
the adaptive algorithm from Section \ref{s4.1} as the key
ingredient, that yields a quantile up to a user-specified error tolerance,
see Section \ref{s5.1}. Numerical experiments, which show
the superiority of the new algorithm over the standard
approach, are presented in Section \ref{s5.2}.

\section{The Statistical Problem: Change-Point Test}\label{sec:CPT}

We are interested in detecting a structural change, specifically at most
one change (AMOC), in a time series model, and for illustration we
consider the following most simple model with a possible mean change. The
model, which is one of the earliest change-point models analyzed,
is given by
\[
\xi_i =
\begin{cases}
\e_i, &\text{if $i\le \cp$},\\
d +\e_i, &\text{if $i> \cp$},\\
\end{cases}
\]
for $i=1,\dots,n$, where $n,m \in \N$ with $n \geq 2$ and
\[
1 \leq \cp = \cp(n) \le n,
\]
and where $d \in \R$ with $d \neq 0$,
see \citet*{Page57}.
The residuals $\e_i$ are assumed to be iid, each
centered with finite second moment $\sigma^2 > 0$, which is assumed
to be known for simplicity. If $m < n$, then a structural
change is present, and $m$ is called the change-point.
A test is constructed for
\[
H_0\colon m=n, \qquad\qquad H_1\colon m<n.
\]
Based on the quasi likelihood ratio test,
the weighted CUSUM statistic
\[
T_n(w) = \max_{1\le k<n}
w(k/n) \cdot \frac{|T_{k,n}|}{\sqrt{n}}
\]
with
\[
T_{k,n} = \sum^k_{i=1}\xi_i-\frac{k}{n}\sum^n_{i=1}\xi_i
\]
and with suitable weight functions $w : \left]0,1\right[ \to
\left[0,\infty\right[$ has been derived. We add that under $H_0$
(no change)
\[
T_{k,n} =
\sum^k_{i=1}\e_i-\frac{k}{n}\sum^n_{i=1}\e_i =
n \cdot \left(\frac{k}{n} \cdot \left(1-\frac{k}{n}\right)\right) \cdot
\left( \frac{1}{k} \sum_{i=1}^k \e_i - \frac{1}{n-k}
\sum_{i=k+1}^n \e_i \right).
\]
See \citet*[Thm.~2.1.2]{book:CsorgoHorvath} for the
following result.

\begin{thm}[Darling-Erd\"os Theorem]\label{t1}
Let
\[
w(t) = (t \cdot (1-t))^{-1/2}
\]
for $0 < t < 1$, and assume that $\E(|\e_i|^{2+\delta}) < \infty$
for some $\delta > 0$. Under $H_0$ we have
\[
\lim_{n \to \infty} P(\{ T_n(w) \leq c_{n}(\alpha)\}) =
1-\alpha
\]
for $0 < \alpha < 1$ and
\[
c_{n}(\alpha) =
\frac{\sigma}{a(\log n)}
\left(-\log \left(-\tfrac{1}{2} \log(1-\alpha)\right)+
b(\log n)\right) ,
\]
with
\[
a(x)=\sqrt{2\log x}
\qquad\text{and}\qquad
b(x)=2\log x +\tfrac{1}{2}\log \log x - \tfrac{1}{2}\log
\pi.
\]
\end{thm}

Theorem~\ref{t1} immediately yields an asymptotic level $\alpha$
test, see Remark~\ref{r2}. However, for small sample sizes
$n$ the convergence in the Darling-Erd\"os
Theorem often leads to level distortion,
see \cite{dissertationKirch}.
To overcome this problem
modifications of the weight function $w$ are considered,
see, e.g., \citet*{CsorgoHorvath88}.
In this paper we study weight functions $w_{\eta,\gamma}$ of the form
\begin{equation}\label{eq:weightfunction}
w_{\eta,\gamma}(t) = 1_{\left]\eta,1-\eta\right[}(t)
\cdot (t \cdot (1-t))^{-\gamma}
\end{equation}
for $0 < t < 1$, where
\[
0 \leq \eta < 1/2, \qquad\qquad 0 \leq \gamma \leq 1/2.
\]
Observe that Theorem~\ref{t1} deals with the case $(\eta,\gamma)=(0,1/2)$.

For any real-valued stochastic process $X=(X(t))_{t \in {]0,1[}}$
we put
\[
S(X) = \sup_{0 < t < 1} X(t).
\]
Furthermore, $B=(B(t))_{t \in [0,1]}$ denotes a standard Brownian
bridge on the unit interval.
The process $|B| = (|B(t)|)_{t \in [0,1]}$ is called a reflecting
Brownian bridge.
The following theorem is a consequence of a general result
from \citet*{incoll:KirchTadjuidje16} and \citet*{Dis:Schwaar},
who study general weight functions under suitable regularity conditions.

\begin{thm}\label{t2}
Let $w_{\eta,\gamma}$ be given by \eqref{eq:weightfunction}
with $(\eta,\gamma) \neq (0,1/2)$.
Under $H_0$ we have
\[
\frac{1}{\sigma} \cdot T_n(w_{\eta,\gamma})\dconv S(w_{\eta,\gamma} |B|).
\]
\end{thm}

Basically, Theorem~\ref{t2} yields an asymptotic level $\alpha$ test.
However,
for application the quantiles of the supremum $S(w_{\eta,\gamma}|B|)$ of
the weighted reflecting Brownian bridge $w_{\eta,\gamma}|B|$
are needed,
i.e., they have to be known analytically or to be easily
computed numerically, see Remark~\ref{r2}.
In \citet*{Schwaar20} {besides data driven weighted change-point estimators, data driven weighted change-point tests are considered, where the parameter $\gamma$ is replaced by an estimator and $\eta=0$}.
Especially then knowledge
about the quantiles for the weighted test statistic with non-extreme
values is required.

For completeness we add that $S(w_{0,1/2}|B|) = \infty$ with
probability one, which follows from the law of the iterated logarithm,
and hence Theorem~\ref{t2} implies
$T_n(w_{0,1/2}) \pconv \infty$.

\begin{rem}\label{r1}
Consider the extremal cases $\gamma = 0$ and $\gamma = 1/2$.
For $(\eta,\gamma) = (0,0)$ the series representation
of the c.d.f.\ of the Kolmogorov distribution may be used to
compute the quantiles.
More generally, for $\gamma=0$ and $0 \leq \eta < 1/2$
a series representation of the c.d.f.\ of the supremum of
a reflecting Brownian motion for given initial and terminal value,
see \citet*[Eqn.~(3.1.1.1.8)]{borodin_salminen_1996},
may be used.

For $\gamma = 1/2$ and $0 < \eta < 1/2$ \citet*{DeLong81} has
studied the distribution of $S(w_{\eta,1/2}|B|)$ and related
quantities, see also \citet*{MR777516}.
The Mellin transform of a corresponding hitting time for a standard
Brownian motion has been determined, and values of the c.d.f. have been
obtained via numerical inversion.
\end{rem}

To the best knowledge of the authors, no (semi-)analytic way to
determine the quantiles is known beyond the extremal cases, i.e.,
for $0 < \gamma < 1/2$ and $0 \leq \eta < 1/2$,
cf.\ \citet*[p.~76]{salminen_yor_2011}.

\begin{rem}\label{r2}
Let $c_{\eta,\gamma,n}>0$.
Consider the test that rejects $H_0$ if and only
if
\[
\frac{|T_{k,n}|}{\sqrt{n}} > c_{\eta,\gamma,n} \cdot
\left(\frac{k}{n} \cdot \left(1-\frac{k}{n}\right)\right)^{\gamma}
\]
for some $k \in \N$ with
\[
\eta \cdot n < k < (1-\eta) \cdot n.
\]
To obtain an asymptotic level $\alpha$ test we proceed as follows.
For $0 \leq \gamma < 1/2$ the critical value
$c_{\eta,\gamma} = c_{\eta,\gamma,n}$ is determined by
\[
P(\{S(w_{\eta,\gamma}|B|) \leq c_{\eta,\gamma}/\sigma \}) = 1 - \alpha,
\]
see Theorem~\ref{t2}, which may be solved semi-analytically if
$\gamma = 0$ and numerically, using the algorithm presented in
Section~\ref{s5},
if $0 < \gamma < 1/2$. For $\gamma=1/2$ we may employ
Theorem~\ref{t1} to determine $c_{\eta,\gamma,n}$.
\end{rem}

\begin{exmp}\label{ex1}
We illustrate the role of the parameter $\gamma$ in the case
$\eta = 0$. Let $\sigma=1$ and $\alpha = 0.05$.
In Figure~\ref{fig:weight_function}
the threshold function
\[
f_{\gamma,n} (t) = c_{0,\gamma,n} \cdot (t \cdot (1-t))^\gamma
\]
is presented for $\gamma=0,\, 0.25,\, 0.45$, and for
$\gamma = 0.5$ with $n=10^2,\, 10^3$.
The critical values are given numerically as follows.
Remark~\ref{r1} yields
$c_{0,0.0} = 1.358$,
and the {adaptive} algorithm $Q^\adap_\varepsilon(w,q)$ according to
Section~\ref{s5} with $\varepsilon = 10^{-2}$ yields
$c_{0,0.25} = 1.99$
and
$c_{0,0.45} = 2.91$.
Furthermore,
$c_{0,0.5,10^2} \approx 3.241$ and
$c_{0,0.5,10^3} \approx 3.353$ due to Theorem~\ref{t1}.

%

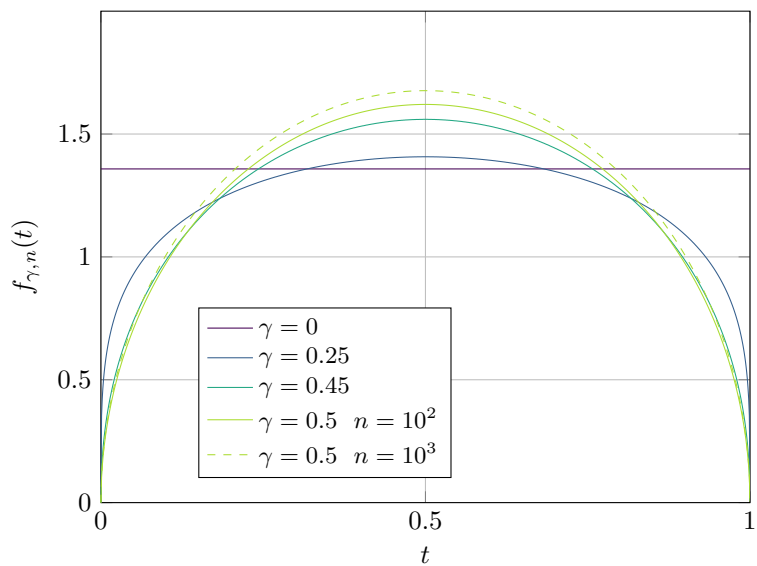
\begin{figure}[htb]
	\centering
	\begin{tikzpicture}
		\begin{axis}[
				xmin = 0, xmax = 1, ymin = 0, ymax = 2,
				xtick={0,0.5,1}, ytick={0,0.5,1,1.5},
				xlabel = {$t$}, ylabel = {$f_{\gamma,n}(t)$},
				legend style = {font=\small,at={(0.15,0.05)},anchor=south west}, legend cell align = left,
				width = 0.8\textwidth,
				height = 0.64\textwidth,
				grid = major,
 				colormap name = {viridis}
			]
			\addplot[no markers, solid, color of colormap={0}, domain=0:1]{1.358};
			\addplot[no markers, solid, color of colormap={300}, domain=0:1]{(\x*(1-\x))^(0.25)*1.99};
			\addplot[no markers, solid, color of colormap={600}, domain=0:1]{(\x*(1-\x))^(0.45)*2.91};
			\addplot[no markers, solid, color of colormap={875}, domain=0:1]{sqrt(x*(1-x))*3.241};
			\addplot[no markers, dashed, color of colormap={875}, domain=0:1]{sqrt(x*(1-x))*3.353};
			\legend{$\gamma=0$, $\gamma=0.25$, $\gamma=0.45$, $\gamma=0.5\ \ n=10^2$, $\gamma=0.5\ \ n=10^3$}
		\end{axis}
	\end{tikzpicture}
	\caption{Threshold function $f_{\gamma,n}$ from Example \ref{ex1}
		for $\eta=0$, $\sigma=1$, and $\alpha = 0.05$.}
	\label{fig:weight_function}
\end{figure}

We observe from Figure~\ref{fig:weight_function}
that the power of the test with a weight function having
$\gamma$ close to $1/2$ is higher for early and late changes.
In the case of a change in the middle, the test
with $\gamma=0$ has a higher power than the one with $\gamma$ close to
$1/2$.
\end{exmp}

\begin{rem}\label{r3}
The construction of change-point tests in more complicated time
series models,
e.g., with a serial dependence, may also be based on CUSUM
statistics and quantiles of $S(X)$ for suitable processes $X$.
See \citet*{AueHorvath13} for models that lead
to $X=w |B|$ with weight functions $w$,
or to $X = \sum_{j=1}^\ell B_j^2$ with independent standard Brownian
bridges $B_j$.
More generally, see \citet*{Aue_et_al_2018} for functional time
series models that lead to $X=\sum_{j=1}^\infty \lambda_j B_j^2$ with
non-negative scalars $\lambda_j$.
Our approach in Section~\ref{s5} can easily be adapted to these
classes of processes.
\end{rem}

\section{Approximation of the Supremum of a Brownian Motion}%
\label{poweradap}

In this section we discuss the strong (or pathwise)
approximation of the supremum $S(W)$
of a standard Brownian motion $W=(W(t))_{t\in [0,1]}$ on the unit
interval. We consider algorithms $A$ that evaluate $W$ at a finite
number of points $t_{k} \in \left]0,1\right]$ and
approximate $S(W)$ by the
discrete maximum of $W$ at these points. The error $e(A)$ of any such
measurable algorithm $A$ is defined by
\[
e(A) = \E \left(|S(W) - A(W)|\right).
\]
We recall known results that demonstrate that
suitable adaptive algorithms, i.e., algorithms that sequentially
evaluate any trajectory of $W$, are far superior to non-adaptive
algorithms, i.e., algorithms that are based on a
fixed, a priori given discretization of $\left]0,1\right]$.

At first we consider the class of all non-adaptive algorithms that use $n$
evaluations of $W$. The following result is due to \citet*{MR1085383}.

\begin{thm}\label{t3}
There exist constants $c_1,c_2 > 0$ with the following properties
for every $n \in \N$.
The algorithm $A_n^\equi$ given by
\[
A_n^\equi (W) = \max_{k=1,\dots,n} W(k/n)
\]
satisfies
\begin{equation}\label{g1}
e(A_n^\equi) \leq c_1 \cdot n^{-1/2}.
\end{equation}
For every choice of $t_1, \dots, t_n \in
\left]0,1\right]$ the algorithm $A^\non_n$ given by
\[
A^\non_n(W) = \max_{k=1,\dots,n} W(t_{k})
\]
satisfies
\begin{equation}\label{g2}
e(A^\non_n) \geq c_2 \cdot n^{-1/2}.
\end{equation}
\end{thm}

For the purpose of this paper the crucial part of
Theorem~\ref{t3} is the
lower bound \eqref{g2}, which says that non-adaptive algorithms achieve at
most the order of convergence $1/2$. This lower bound is sharp, up to
constants, as we have a matching upper bound \eqref{g1}, which is already
achieved by an equidistant discretization.

\begin{rem}\label{r4}
We conjecture that the lower bound \eqref{g2} is also valid for algorithms
$A^\non_n$ of the form $A^\non_n(W)=\zeta(W(t_1),\dots, W(t_n))$,
where $\zeta\colon \R^n\to\R$ is any measurable mapping.
The conjecture is true at least if $t_{k}=k/n$, see
\citet*[Prop.~2.1 and proof of Thm.~3.3]{MR3744680}, or if the
$L_2$-error of $A^\non_n$ is considered, instead of the $L_1$-error
$e(A^\non_n)$, see \citet*[beginning of proof of Thm.~2.1]{Calvin2004}.

The asymptotic distribution of the error $S(W) - A_n^\equi
(W)$ has been derived in \citet*[Thm.~1]{MR1384357}.
\end{rem}

An adaptive algorithm $A^\adap_n$ that uses $n$ evaluations of the
Brownian motion $W$ is formally defined by a point
$t_1 \in \left]0,1\right]$ and Borel-measurable mappings
\[
\chi_k \colon \R^{k-1} \to \left]0,1\right]
\]
for $k=2,\dots,n$. Iteratively the algorithm computes
$y_1 = W(t_1)$ and
\[
y_k = W (\chi_k(y_1,\dots,y_{k-1}))
\]
for $k=2,\dots,n$, and it yields the output
\[
A^\adap_n (W) = \max_{k=1,\dots,n} y_{k}.
\]
In this way the $k$-th evaluation site
$\chi_k(y_1,\dots,y_{k-1})$ may depend on the previously
obtained values $y_1,\dots,y_{k-1}$ (and the corresponding evaluation
sites).
Of course, the non-adaptive algorithms, which are considered in
Theorem~\ref{t3}, correspond to the particular case of constant mappings
$\chi_k$.

The following result is due to \citet*{MR3605752}.

\begin{thm}\label{t4}
There exists a sequence of adaptive algorithms $A_n^\adap$ with the
following property. For every $\rho > 0$ there exists a constant
$c > 0$ such that for every $n \in \N$
\[
e(A^\adap_n) \leq c \cdot n^{-\rho}.
\]
\end{thm}

We refer to \citet*{MR3605752} for the construction of the algorithms
$A^\adap_n$ that are considered in Theorem~\ref{t4};
see also Section~\ref{s3} for basic ideas.

According to Theorem~\ref{t4} suitable adaptive algorithms achieve,
roughly speaking, the polynomial order of convergence $\infty$.
Combining Theorems~\ref{t3} and \ref{t4} we see that adaptive algorithms
strikingly outperform all non-adaptive algorithms for the
strong approximation of the supremum $S(W)$ of a Brownian motion.

Of course, Theorems~\ref{t3} and \ref{t4} are irrelevant
for the computation of quantiles of $S(W)$, since the distribution
of $S(W)$ is known explicitly. The theorems strongly suggest, however, that
adaptive algorithms should be considered for quantile computation
if (semi-)analytic methods are not available. The latter holds true for
the processes $w_{\eta,\gamma} |B|$ with $\gamma \not \in \{0,1/2\}$,
see Section~\ref{sec:CPT}.

\section{Approximation of the Supremum
of a Weighted Reflecting Brownian Bridge}\label{s3}

For notational convenience we put
\[
w = w_{\eta,\gamma},
\]
where $(\eta,\gamma) \neq (0,1/2)$.

\subsection{The Adaptive Algorithm}\label{s4.1}

In this section we present an adaptive algorithm for the strong
approximation of the supremum $S(w |B|)$ of the weighted reflecting
Brownian bridge $w|B|$. This algorithm is a modification of
the algorithm constructed in \citet*{MR3605752}, which achieves the error
bound in Theorem~\ref{t4} for the strong approximation of the supremum
$S(W)$ of a standard Brownian motion $W$.

Both of these adaptive algorithms are greedy algorithms,
and the basic idea in the construction is as follows.
After the $k$-th step the algorithm
has computed a partition of the interval $[0,1]$ into
$k$ subintervals together with the values of the underlying stochastic
process at the boundary points of all these intervals.
A score value is available for each subinterval,
and the subinterval with the highest score will
be split at the midpoint. Ideally,
the score value should be the conditional probability that
the corresponding subinterval contains a global maximizer
of $w |B|$. Reasonable substitutes for these conditional probabilities
are needed in the computation.

In the sequel we present the algorithm for the strong approximation of
$S(w|B|)$ in detail. Based on the weight function $w$ we assign a weight
$v(\ell,r)$ to any interval $[\ell,r] \subseteq [0,1]$ with positive
length in the following way. Let
\[
c = \frac{\ell+r}{2}.
\]
For a weight function $w$ that is positive and differentiable
with a `small' derivative everywhere on $\left]0,1\right[$
it is reasonable to take $v(\ell,r) =w(c)$. Since these conditions are not
met for $w = w_{\eta,\gamma}$, except for the trivial case
$(\eta,\gamma)=0$, we proceed differently.
To avoid a too small score value we take
\[
v(\ell,r)
=\begin{cases}
0,  & \text{if $[\ell,r]\subseteq [0,\eta]\cup [1-\eta,1]$},\\
(c \cdot (1-c))^{-\gamma}, & \text{otherwise.}
\end{cases}
\]
In fact, observe that $v(\ell,r) \neq w(c)$ if and only if
$c \leq \eta < r$ or $\ell < 1-\eta \leq c$. If $v(\ell,r) \neq
w(c)$, then $w(c) = 0$ while $v(\ell,r)$ may
potentially be very large.

Suppose that $x = B(\ell)$ and $y = B(r)$ are known, while
no values of $B$ are known inside of $\left]\ell,r\right[$.
It is reasonable to compare $v(\ell,r) \cdot B(c)$ with a certain
threshold $m$, e.g., with the largest value of $w |B|$ known so far,
under the conditional distribution of $B(c)$. The latter is the normal
distribution with mean $(x+y)/2$ and variance $(r-\ell)/4$.
More precisely, we define the score function
\[
\varphi \colon D \times \R^2 \times \left[0,\infty\right[ \to
\left[0,\infty\right[
\]
by
\[
\varphi(\ell,r,x,y,m)
=
\E \left( (v(\ell,r)\cdot Z-m)^+\right)
+
\E \left( (v(\ell,r)\cdot Z+m)^-\right),
\]
where
\[
D = \{ (\ell,r) \in [0,1]^2 : \ell < r \}
\]
and
\[
Z \sim N((x+y)/2, (r-\ell)/4).
\]

\begin{rem}\label{r5}
The score function is easily computed as follows.
Let $\Phi$ denote the distribution function of $Y\sim N(0,1)$, and
let $a \in \R$. Put
\[
\psi (a) = \Phi^\prime(a) + a \cdot \Phi(a).
\]
Since $\E \left( (Y+a)^+ \right) = \psi(a)$, we obtain
\begin{align*}
&\varphi(\ell,r,x,y, m)\\
&=
v(\ell,r) \cdot \frac{\sqrt{r-\ell}}{2}
\cdot \left(
\psi\left(\frac{x+y-2m/v(\ell,r)}{\sqrt{r-\ell}}\right)
+
\psi\left(-\frac{x+y+2m/v(\ell,r)}{\sqrt{r-\ell}}\right)
\right)
\end{align*}
if $v(\ell,r)>0$. Otherwise
$\varphi(\ell,r,x,y,m) = 0$, since $m \geq 0$.

To avoid rounding errors in the evaluation of $\psi$ one may use
\[
\lim_{a\to\infty} \frac{\psi(a)}{a}=1
\]
and
\[
\lim_{a \to -\infty}
\frac{\psi(a)}{a^{-2} \cdot \Phi^\prime(a)} = 1.
\]
We use this asymptotic behavior for $|a|>3$, i.e., we replace $\psi$ by
\begin{align*}
\widetilde{\psi}(a)
=
\begin{cases}
a,  & \text{if $a> 3$},\\
a^{-2} \cdot \Phi^\prime(a), & \text{if $a<-3$},\\
\psi(a), & \text{otherwise},
\end{cases}
\end{align*}
in the computation of $\varphi(\ell,r,x,y, m)$.
\end{rem}

We are ready to define the adaptive algorithm, which will be
denoted by $A^\adap_n(w,\cdot)$.  The algorithm will sequentially
evaluate any trajectory of $B$,
and the relevant information about the corresponding partition of $[0,1]$
after $k$ steps is represented by a set $\I_k$ as follows.
There are $k$ elements in $\I_k$, which correspond
to the $k$ subintervals in this partition. More precisely,
$(\ell,r,x,y,s) \in \I_k$ represents a subinterval $[\ell,r]$ with
boundary values $x=B(\ell)$ and $y=B(r)$ and with score value $s$.
Furthermore, $m_k \geq 0$ denotes the discrete maximum of $w |B|$ after $k$
steps.
In the first step we put
\begin{align*}
m_1=0, \qquad \qquad
\I_1=\{(0,1,0,0,\varphi(0,1,0,0,m_1))\}.
\end{align*}
In the $k$-th step with $2 \leq k \leq n$ we
choose any $I = (\ell,r,x,y,s)\in \I_{k-1}$ with the maximal value of $s$
among all elements of $\I_{k-1}$,
i.e., with the largest score value (this choice needs not to be unique),
and we evaluate $B$ at the midpoint of the corresponding
subinterval, i.e., we compute
\begin{align*}
z=B(c).
\end{align*}
The new discrete maximum of $w |B|$ is given by
\[
m_{k} = \max\left(m_{k-1}, w(c)\cdot |z|\right),
\]
and the new partition is represented by
\[
\I_{k} =\left(\I_{k-1}\setminus\{I\}\right)\cup \{I_1,I_2\}
\]
with
\[
I_1 = (\ell,c,x,z, \varphi(\ell,c,x,z, m_{k})), \qquad
I_2 = (c,r,z,y,\varphi(c,r,z,y, m_{k})).
\]
After $n$ steps the adaptive algorithm returns the output
\[
A^\adap_n(w,B)=m_n.
\]

\begin{rem}\label{r6}
We discuss the computational cost to simulate $A^\adap_n(w,B)$.
First of all, we use a max-priority queue for storing the elements of
the sets $\I_1,\dots,\I_n$. This allows to choose
$(\ell,r,x,y,s)\in \I_{k-1}$ with the maximal value of $s$ at a
cost of order $\ln(k)$ in the $k$-th step, see
\citet*[Sec.~6.5]{cormen2009introduction}. Thereafter $B(c)$ may be
simulated at a constant cost, independently of $k$,
under the conditional distribution. By definition, no updates
of the score values are computed for the elements in $\I_k \cap
\I_{k-1}$, although $m_k$ may be larger than $m_{k-1}$.
It follows that the total cost to simulate $A^\adap_n(w,B)$
is of the order $n \ln(n)$.
\end{rem}

\subsection{Numerical Experiments}\label{strongnumeriocs}

We compare the adaptive algorithm $A^\adap_n(w,\cdot)$
and the non-adaptive algorithm $A^\equi_n(w,\cdot)$ given by
\[
A_n^\equi (w,B) = \max_{k=1,\dots,n-1} w(t_{k})
\cdot |B(t_{k})|
\]
with $t_{k} = k/n$, which has been used by, e.g.,
\citet*{EastwoodEastwood98} and \citet*{OraschPouliot04},
and for a similar problem by \citet*{AkashiEtAl18}.

Analogously to Section~\ref{poweradap} we consider the error
\begin{equation}\label{g77}
e(A_n(w,\cdot)) = \E \left(|S(w|B|) - A_n(w,B)|\right)
\end{equation}
for $A_n(w,\cdot) = A_n^\adap(w,\cdot)$ and
$A_n(w,\cdot) = A_n^\equi(w,\cdot)$.
We add that exact values or upper or lower bounds
of the error
are not available for any of these algorithms.
Therefore we determine the error via
a Monte Carlo simulation, where we replace $S(w|B|)$
in \eqref{g77} by
$A^\adap_{n_0}(w,B)$ or $A^\equi_{n_0}(w,B)$,
respectively,
with $n_0$ being $10$ times larger than the largest value of $n$
that is considered in the numerical experiment.
Simultaneously, we determine the average run-time of both
algorithms.
For each value of $n$ we use $10^3$ Monte Carlo replications.

All programs are written in C++,
and the computations are performed on
a single Intel Xeon Gold 6126 processor.
All results are presented together with asymptotic confidence
intervals with confidence level $0.95$.
In the numerical experiments we consider $\eta = 0$ and
$\gamma=0.25$ or $\gamma=0.45$, cf.\ Example \ref{ex1}.

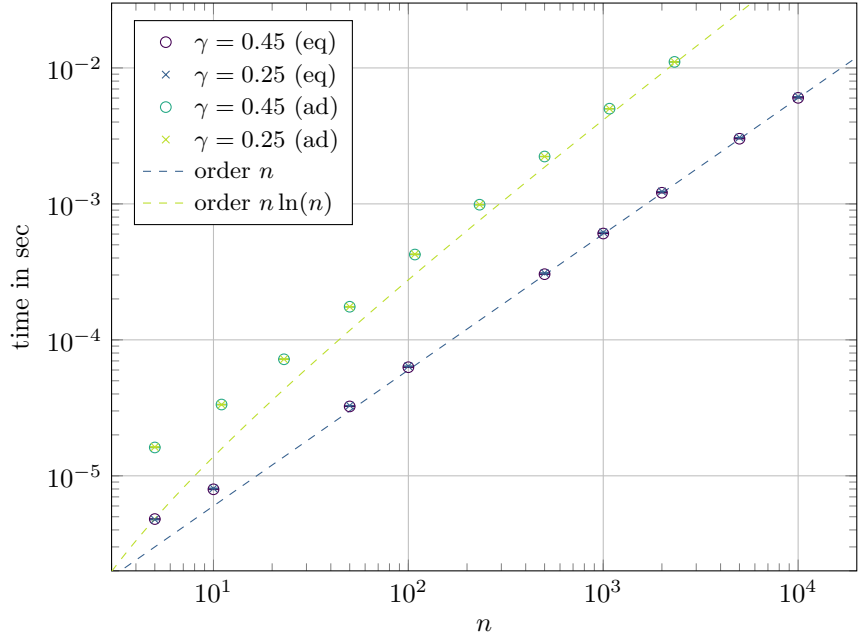
\begin{figure}
	\centering
	\begin{tikzpicture}
		\begin{loglogaxis}[
				xmin = 3, xmax = 2*10^(4), ymin = 2*10^(-6), ymax = 3*10^(-2),
				xlabel = {$n$}, ylabel = {time in sec},
				legend style = {font=\small}, legend pos = north west, legend cell align = left,
				width = 0.9\textwidth,
				height = 0.72\textwidth,
				grid = major,
				colormap name = viridis
			]
			\addplot[only marks, mark=o, mark size=2pt, color of colormap={0}]
				plot[error bars/.cd, y dir=both, y explicit]
				table[x=n, y=time, y error=time_delta]
				{
				n error time error_delta time_delta
				5 0.845162302674577 4.80869500000000e-06 0.0271286729568607 2.49378395839871e-08
				10 0.635046672203047 7.95890100000001e-06 0.0228855549383234 3.83300360107666e-08
				50 0.332916705644544 3.23810460000000e-05 0.0165797965771209 5.79553865264049e-08
				100 0.257372844230698 6.29159870000000e-05 0.0135070752079162 9.03367613793682e-08
				500 0.138527134498256 0.000304081914000000 0.00962827105671228 1.51402982503953e-07
				1000 0.0981835797671943 0.000605440564000000 0.00764813019389529 1.90494492276949e-07
				2000 0.0763831526809399 0.00120731175000000 0.00629755246999791 2.32894418394823e-07
				5000 0.0462462344613881 0.00301398554200000 0.00405346862550486 4.24059012168403e-07
				10000 0.0304972559878189 0.00602490878000000 0.00355324373841582 6.84499651744857e-07
				};
			\addplot[only marks, mark=x, mark size=2pt, color of colormap={300}]
				plot[error bars/.cd, y dir=both, y explicit]
				table[x=n, y=time, y error=time_delta]
				{
				n error time error_delta time_delta
				5 0.429859249852026 4.76886700000000e-06 0.0130934306177733 2.80243123192506e-08
				10 0.294507623942235 8.08176399999999e-06 0.00957162492302653 4.08223994864251e-08
				50 0.128986501934025 3.28726380000000e-05 0.00424117114357662 6.28673391625472e-08
				100 0.0900469146283198 6.40675950000000e-05 0.00296579664358858 7.94614506271395e-08
				500 0.0392517679653400 0.000310431944000000 0.00144358402635864 1.44284207254509e-07
				1000 0.0264734082476520 0.000618128421000001 0.00100079995830678 1.86651446280843e-07
				2000 0.0186691886738727 0.00123388321400000 0.000736302864600501 2.41831032357930e-07
				5000 0.0105051317355712 0.00307788876000000 0.000452729801528795 3.01844101252326e-07
				10000 0.00657070915793112 0.00615493150999999 0.000311692157081604 7.19239127556640e-07
				};
			\addplot[only marks, mark=o, mark size=2pt, color of colormap={600}]
				plot[error bars/.cd, y dir=both, y explicit]
				table[x=n, y=time, y error=time_delta]
				{
				n error time error_delta time_delta
				5 0.841602381690364 1.61623610000000e-05 0.0280556188093850 5.71521690241535e-08
				11 0.503065729375378 3.34497799999999e-05 0.0213404446118101 1.05782589404933e-07
				23 0.255763114604254 7.19538230000000e-05 0.0155442108061990 2.57663295326493e-07
				50 0.108889828970193 0.000175230805000000 0.0101918652389972 6.01012858653012e-07
				108 0.0307267520989333 0.000424115865000000 0.00540563939055149 1.19155860007369e-06
				232 0.00485878440089793 0.000986308653000000 0.00224094633216035 1.89728779420480e-06
				500 3.47795293175937e-05 0.00223082075300000 3.18813642668982e-05 3.08387221265659e-06
				1077 5.87700415977110e-09 0.00501559507500000 4.72132865141030e-09 5.21948033014351e-06
				2321 9.24949006275710e-15 0.0110753775710000 5.37766665352089e-15 8.66189062549133e-06
				};
			\addplot[only marks, mark=x, mark size=2pt, color of colormap={900}]
				plot[error bars/.cd, y dir=both, y explicit]
				table[x=n, y=time, y error=time_delta]
				{
				n error time error_delta time_delta
				5 0.425466181054310 1.62904310000000e-05 0.0138689407860020 6.93122205518823e-08
				11 0.204018197263233 3.33828820000000e-05 0.00822364117501181 9.21912729660278e-08
				23 0.0864910123905192 7.20632840000000e-05 0.00530668705312227 2.87264283115858e-07
				50 0.0253835368383637 0.000175225671000000 0.00248145495581894 7.32440229175159e-07
				108 0.00430764972725310 0.000426489420000000 0.000849395911197680 1.36225184217530e-06
				232 0.000310839442108387 0.000987867712000000 0.000101226920088781 2.21664629208564e-06
				500 9.91341050757244e-06 0.00223068009700000 1.60811331778122e-05 3.55144059035064e-06
				1077 8.28192554314100e-09 0.00501369936200000 1.41681043109157e-08 5.33465041723152e-06
				2321 1.14097620240727e-15 0.0110752219150000 5.84194837806379e-16 8.40858594615147e-06
				};
			\addplot[no markers, dashed, color of colormap={300},   domain=3:2*10^(4)]{6*10^(-7)*x};
			\addplot[no markers, dashed, color of colormap={900}, domain=3:2*10^(4)]{6*10^(-7)*x*ln(x)};
			\legend{$\gamma=0.45$ (eq),
				$\gamma=0.25$ (eq),
				$\gamma=0.45$ (ad),
				$\gamma=0.25$ (ad),
				order $n$, order $n\ln(n)$}
		\end{loglogaxis}
	\end{tikzpicture}
	\caption{Average run-time vs.\ number $n$ of discretization points
		for the strong approximation of $S(w|B|)$.}
	\label{fig:strong:2}
\end{figure}

In Figure~\ref{fig:strong:2} we relate the average run-time
to the number $n$ of discretization points.
For $n \leq 10^4$ the computational overhead due to
adaption is moderate, as the non-adaptive algorithm is at most 10 times
faster than the adaptive algorithm in this range. Furthermore,
the average run-times for both algorithms
are in line with the worst case behavior, namely,
order $n \ln (n)$ for $A_n^\adap(w,\cdot)$, see Remark \ref{r6},
and order $n$ for $A_n^\equi(w,\cdot)$.
{Finally, we see that the average run-time does not depend on
$\gamma$.}

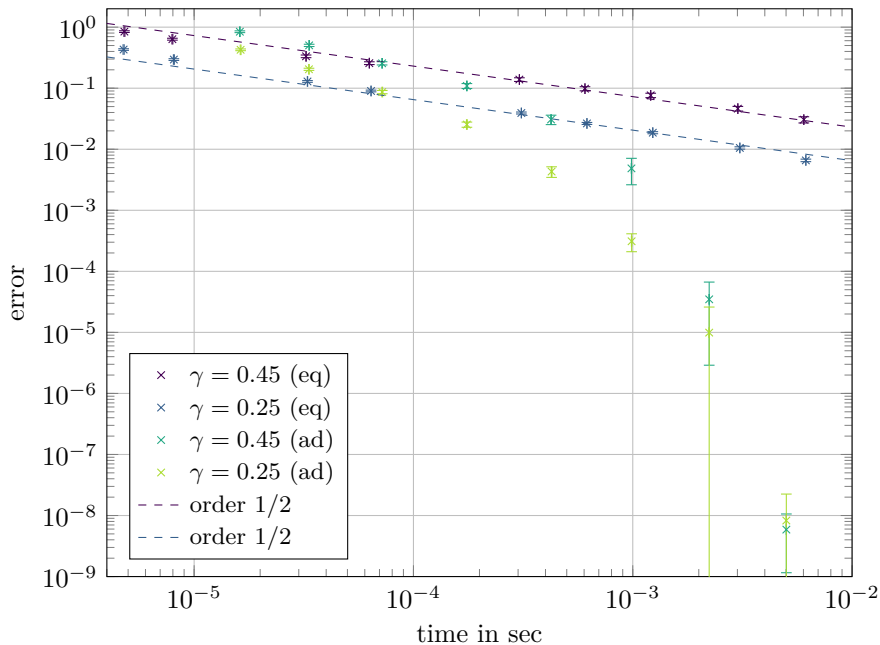
\begin{figure}[htb]
	\centering
	\begin{tikzpicture}
		\begin{loglogaxis}[
				xmin = 4*10^(-6), xmax = 10^(-2), ymin = 10^(-9), ymax = 2*10^(0),
				xlabel = {time in sec}, ylabel = {error},
				legend style = {font=\small}, legend pos = south west, legend cell align = left,
				width = 0.9\textwidth,
				height = 0.72\textwidth,
				grid = major,
				colormap name = viridis
			]
			\addplot[only marks, mark=x, mark size=2pt, color of colormap={0}]
				plot[error bars/.cd, x dir=both, x explicit, y dir=both, y explicit]
				table[x=time, y=error, x error=time_delta, y error=error_delta]
				{
				n error time error_delta time_delta
				5 0.845162302674577 4.80869500000000e-06 0.0271286729568607 2.49378395839871e-08
				10 0.635046672203047 7.95890100000001e-06 0.0228855549383234 3.83300360107666e-08
				50 0.332916705644544 3.23810460000000e-05 0.0165797965771209 5.79553865264049e-08
				100 0.257372844230698 6.29159870000000e-05 0.0135070752079162 9.03367613793682e-08
				500 0.138527134498256 0.000304081914000000 0.00962827105671228 1.51402982503953e-07
				1000 0.0981835797671943 0.000605440564000000 0.00764813019389529 1.90494492276949e-07
				2000 0.0763831526809399 0.00120731175000000 0.00629755246999791 2.32894418394823e-07
				5000 0.0462462344613881 0.00301398554200000 0.00405346862550486 4.24059012168403e-07
				10000 0.0304972559878189 0.00602490878000000 0.00355324373841582 6.84499651744857e-07
				};
			\addplot[only marks, mark=x, mark size=2pt, color of colormap={300}]
				plot[error bars/.cd, x dir=both, x explicit, y dir=both, y explicit]
				table[x=time, y=error, x error=time_delta, y error=error_delta]
				{
				n error time error_delta time_delta
				5 0.429859249852026 4.76886700000000e-06 0.0130934306177733 2.80243123192506e-08
				10 0.294507623942235 8.08176399999999e-06 0.00957162492302653 4.08223994864251e-08
				50 0.128986501934025 3.28726380000000e-05 0.00424117114357662 6.28673391625472e-08
				100 0.0900469146283198 6.40675950000000e-05 0.00296579664358858 7.94614506271395e-08
				500 0.0392517679653400 0.000310431944000000 0.00144358402635864 1.44284207254509e-07
				1000 0.0264734082476520 0.000618128421000001 0.00100079995830678 1.86651446280843e-07
				2000 0.0186691886738727 0.00123388321400000 0.000736302864600501 2.41831032357930e-07
				5000 0.0105051317355712 0.00307788876000000 0.000452729801528795 3.01844101252326e-07
				10000 0.00657070915793112 0.00615493150999999 0.000311692157081604 7.19239127556640e-07
				};
			\addplot[only marks, mark=x, mark size=2pt, color of colormap={600}]
				plot[error bars/.cd, x dir=both, x explicit, y dir=both, y explicit]
				table[x=time, y=error, x error=time_delta, y error=error_delta]
				{
				n error time error_delta time_delta
				5 0.841602381690364 1.61623610000000e-05 0.0280556188093850 5.71521690241535e-08
				11 0.503065729375378 3.34497799999999e-05 0.0213404446118101 1.05782589404933e-07
				23 0.255763114604254 7.19538230000000e-05 0.0155442108061990 2.57663295326493e-07
				50 0.108889828970193 0.000175230805000000 0.0101918652389972 6.01012858653012e-07
				108 0.0307267520989333 0.000424115865000000 0.00540563939055149 1.19155860007369e-06
				232 0.00485878440089793 0.000986308653000000 0.00224094633216035 1.89728779420480e-06
				500 3.47795293175937e-05 0.00223082075300000 3.18813642668982e-05 3.08387221265659e-06
				1077 5.87700415977110e-09 0.00501559507500000 4.72132865141030e-09 5.21948033014351e-06
				2321 9.24949006275710e-15 0.0110753775710000 5.37766665352089e-15 8.66189062549133e-06
				};
			\addplot[only marks, mark=x, mark size=2pt, color of colormap={875}]
				plot[error bars/.cd, x dir=both, x explicit, y dir=both, y explicit]
				table[x=time, y=error, x error=time_delta,
					y error plus=error_delta, y error minus=error_delta_cutoff]
				{
				n error time error_delta time_delta error_delta_cutoff
				5 0.425466181054310 1.62904310000000e-05 0.0138689407860020 6.93122205518823e-08 0.0138689407860020
				11 0.204018197263233 3.33828820000000e-05 0.00822364117501181 9.21912729660278e-08 0.00822364117501181
				23 0.0864910123905192 7.20632840000000e-05 0.00530668705312227 2.87264283115858e-07 0.00530668705312227
				50 0.0253835368383637 0.000175225671000000 0.00248145495581894 7.32440229175159e-07 0.00248145495581894
				108 0.00430764972725310 0.000426489420000000 0.000849395911197680 1.36225184217530e-06 0.000849395911197680
				232 0.000310839442108387 0.000987867712000000 0.000101226920088781 2.21664629208564e-06 0.000101226920088781
				500 9.91341050757244e-06 0.00223068009700000 1.60811331778122e-05 3.55144059035064e-06 9.9134e-06
				1077 8.28192554314100e-09 0.00501369936200000 1.41681043109157e-08 5.33465041723152e-06 8.2819e-09
				2321 1.14097620240727e-15 0.0110752219150000 5.84194837806379e-16 8.40858594615147e-06 5.84194837806379e-16
				};
			\addplot[no markers, dashed, color of colormap={0},   domain=4*10^(-6):10^(-2)]{0.0023*x^(-0.5)};
			\addplot[no markers, dashed, color of colormap={300}, domain=4*10^(-6):10^(-2)]{0.00065*x^(-0.5)};
			\legend{$\gamma=0.45$ (eq),$\gamma=0.25$ (eq),$\gamma=0.45$ (ad),$\gamma=0.25$ (ad),
				order $1/2$, order $1/2$}
		\end{loglogaxis}
	\end{tikzpicture}
	\caption{Error given by \eqref{g77} vs.\ average run-time
		for the strong approximation of $S(w|B|)$.}
	\label{fig:strong:3}
\end{figure}

The relation between the error and the average run-time, which is
most important, is presented in Figure~\ref{fig:strong:3}.
First of all, we observe a polynomial order of convergence $\infty$
for the adaptive algorithm, in contrast to a polynomial
order of convergence of only about $1/2$ for the non-adaptive
algorithm. This is in line with the corresponding theoretical results
for Brownian motion, see Theorem~\ref{t3} and Theorem~\ref{t4}.
We add that the confidence intervals
regarding time are rather small and therefore
not visible.
As to be expected, a stronger singularity of the weight function, i.e.,
a larger value of $\gamma$, deteriorates the speed of convergence
for both algorithms.
For the same average run-time of $10^{-2}$ seconds the adaptive algorithm
achieves an error of less than $10^{-8}$ for both values of $\gamma$,
while the error of the non-adaptive algorithm is about $10^{-2}$.
For quantile computation we may therefore sample almost from the correct
distribution by means of the adaptive algorithm with a reasonable average
run-time, while this is impossible by means of the non-adaptive
algorithm.

\section{Quantile Computation for a Weighted Reflecting Brownian Bridge}
\label{s5}

In this section we present an algorithm
for the computation of the $q$-quantile of $S(w |B|)$.
The inputs of this algorithm, which will be denoted by
$Q^\adap_\e(w, q)$,
are the weight $w$, i.e., $\eta$ and $\gamma$, as well as
$0 < q < 1$ and an error tolerance $\e>0$.
The key ingredient is the algorithm $A^\adap_n(w,\cdot)$ from
Section~\ref{s3}. Our construction is rather ad hoc
and many improvements are possible. We mainly want to demonstrate the
potential of using an adaptive algorithm as the building block
for quantile computation.

\subsection{The Algorithm}\label{s5.1}

The algorithm $Q^\adap_\e(w, q)$ starts with a precomputing step in order
to ideally determine the minimal integer $n_0\in\N$ such that
\[
e(A^\adap_{n_0}(w,\cdot)) \leq \e.
\]
Due to the fast convergence of $A^\adap_{n}(w,B)$ towards the
supremum $S(w|B|)$, which has been observed in the numerical
experiments in Section \ref{s3}, we use $\E(\Delta_n)$ with
\[
\Delta_n=|A^\adap_{2n}(w,B)-A^\adap_n(w,B)|
\]
as an approximation to $e(A^\adap_n(w,\cdot))$.
Moreover, we use a simple Monte Carlo algorithm
$X_n^{(m)}$ with $m$ independent samples of
$\Delta_n$ to approximate the expectation $\E(\Delta_n)$.
In the precomputing step we take
\[
m = 10^3,
\]
and we compute the minimal integer $n_0 \in \N$ of the form
\[
n_0 = 10 \cdot 2^i
\]
with $i \in \N_0$ such that
\[
X^{(m)}_{n_0} \leq \e.
\]
If no such $n_0$ exists, then the output of the algorithm
$Q^\adap_\e(w, q)$ is undefined. We add that
this happens with probability zero if, as we conjecture,
$\Delta_n \pconv 0$.

In the second step we generate a certain number $k_0$ of
independent samples of $A^\adap_{n_0}(w,B)$, which are independent
from the precomputing step, too. The choice of this number is
motivated by the following fact.

\begin{rem}\label{r7}
Consider iid random variables $Y_1, Y_2, \dots$
with a continuous density function $f$. Assume
that $f(F^{-1}(q))>0$ for the $q$-quantile $F^{-1}(q)$ of $Y_1$, and let
$Z_{q,k}$ denote the $\lceil q\cdot k\rceil$-th
order statistic of $Y_1,\dots,Y_k$.
Then
\begin{equation*}
c_q \cdot \sqrt{k} \cdot (Z_{q,k}-F^{-1}(q)) \dconv Z
\end{equation*}
with $Z \sim N(0,1)$ and
\[
c_q = \frac{f(F^{-1}(q))}{\sqrt{q \cdot (1-q)}},
\]
see, e.g., \citet*[Thm.~10.3]{MR1994955}.
\end{rem}

Replacing the unknown constant $c_q$ from Remark \ref{r7} by one, we take
\[
k_0=\lceil \varepsilon^{-2}\rceil
\]
samples of $A_{n_0}^\adap(w,B)$, and the algorithm
$Q^\adap_{\varepsilon}(w, q)$ returns the $\lceil q\cdot k_0\rceil$-th
order statistic of these samples.

The following fact may be used to compute confidence
intervals for the $q$-quantile of $A^\adap_{n_0} (w,B)$,
which yields a quality control for the second step.
Observe, however, that the precomputing step does not yield
a rigorous link to the $q$-quantile of $S(w|B|)$.

\begin{rem}\label{r8}
Consider iid random variables $Y_1, \dots, Y_k$, and let
$Y_{(i)}$ the corresponding $i$-th order statistic.
Furthermore, let $0 < \alpha < 1$, and assume that
$a, b \in \{1,\dots,k\}$ with $a < b$ satisfy
\[
P(\{Z\in \{a,\dots, b-1\}\}) \geq 1-\alpha,
\]
where $Z$ is binomially distributed with parameters $k$ and $q$.
Then $[Y_{(a)},Y_{(b)}]$ is a (conservative) confidence
interval for the $q$-quantile of $Y_1$ with confidence level
$1-\alpha$, see, e.g., \citet*[Sec.~7.1]{MR1994955}.
\end{rem}

\subsection{Numerical Experiments}\label{s5.2}

We compare the algorithm $Q^\adap_{\varepsilon}(w, q)$
and an algorithm $Q^\equi_{\varepsilon}(w, q)$ that is constructed as
$Q^\adap_{\varepsilon}(w, q)$, but instead of
$A_n^\adap (w,\cdot)$ the non-adaptive algorithm $A_n^\equi (w,\cdot)$
is used as the building block. Furthermore, in the precomputing
step of $Q^\equi_{\varepsilon}(w, q)$
we fully use the findings from Figures~\ref{fig:strong:2} and
\ref{fig:strong:3} for free in order to determine the value
of $n_0$. This comes very close to choosing exactly and
at no computational cost the minimal
integer $n_0 \in \N$ such that $e(A_{n_0}^\equi(w,\cdot)) \leq
\varepsilon$, and thus is very much in favor
of $Q^\equi_{\varepsilon}(w, q)$ compared to
of $Q^\adap_{\varepsilon}(w, q)$.

We consider the error
\begin{align}\label{eq:error-quant}
e(Q_\varepsilon(w,q))
=\E\left(\left|F^{-1}(w,q)-Q_\varepsilon(w,q)\right|\right)
\end{align}
for $Q_\varepsilon(w,q)=Q^\adap_{\varepsilon}(w, q)$ and
$Q_\varepsilon(w,q)=Q^\equi_{\varepsilon}(w, q)$,
where $F^{-1}(w,q)$ denotes the $q$-quantile of $S(w |B|)$.

We proceed as in the previous section. The only difference is that
a deterministic quantity, $F^{-1}(w,q)$, instead of a random
variable, $S(w|B|)$, is unknown in the definition of the error.

The error $e(Q_\varepsilon(w,q))$ and the
average run-time are determined via a Monte Carlo simulation, where
we use $10^2$ replications for each value of
$\varepsilon$, and the
results are presented together with asymptotic confidence intervals
as before. We also use the same set of parameters $\eta$
and $\gamma$, the same hardware system, and the same programming
language as before. The value of $q$ is chosen as $q = 0.95$.

In the Monte Carlo simulation we replace $F^{-1}(w,q)$ by a
highly accurate approximation,
namely $2.0008$ for $\gamma=0.25$ and $2.9222$ for $\gamma=0.45$.
The latter is obtained as
the $\lceil q\cdot k_0\rceil$-th order statistic
of $k_0$ samples of $A^\adap_{n_0}(w,B)$, where
$n_0=10^3$ and where $k_0$ is close to $4\cdot 10^8$.
According to the findings from Section \ref{s3}
the distributions of $A^\adap_{n_0}(w,B)$ and $S(w |B|)$ should almost
coincide for this value of $n_0$. Furthermore, using
Remark~\ref{r8} with confidence level $0.99$,
we have obtained the confidence intervals
$[2.0006, 2.0010]$ for $\gamma=0.25$
and $[2.9220,2.9224]$ for $\gamma=0.45$.
We add that this master computation has required a run-time of
about three weeks for each value of $\gamma$.

%
%
%

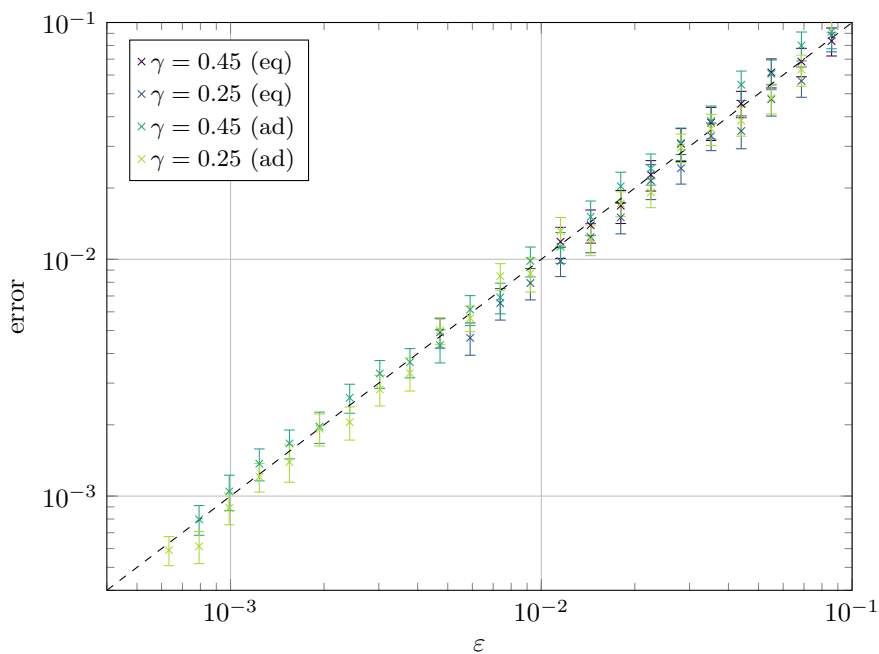
\begin{figure}[htb]
	\centering
	\begin{tikzpicture}
		\begin{loglogaxis}[
				xmin = 0.4*10^(-3), xmax = 10^(-1), ymin = 0.4*10^(-3), ymax = 10^(-1),
				xlabel = {$\varepsilon$}, ylabel = {error},
				legend style = {font=\small}, legend pos = north west, legend cell align = left,
				width = 0.9\textwidth,
				height = 0.72\textwidth,
				grid = major,
				colormap name = viridis
			]
			\addplot[only marks, mark=x, mark size=2pt, color of colormap={0}]
				plot[error bars/.cd, y dir=both, y explicit]
				table[x=eps, y=error, y error=error_delta]
				{
				eps error error_delta
				1 1.67428851344200 0.108600164151742
				0.800000000000000 1.15246924008400 0.105568224230612
				0.640000000000000 0.879000412470000 0.0866898525343202
				0.512000000000000 0.656359359650000 0.0649775619680158
				0.409600000000000 0.392149124490000 0.0461602250370232
				0.327680000000000 0.306268596940000 0.0447158179995266
				0.262144000000000 0.281165813660000 0.0458996766102763
				0.209715200000000 0.196206040070000 0.0287127894558553
				0.167772160000000 0.153873905360000 0.0278707287334233
				0.134217728000000 0.117503663360000 0.0194940202683091
				0.107374182400000 0.0936704797600000 0.0152909661297684
				0.0858993459200001 0.0835105300600000 0.0113085949088017
				0.0687194767360000 0.0683912413300000 0.00938525443405078
				0.0549755813888000 0.0615366029000000 0.00863366371607376
				0.0439804651110400 0.0454179828000000 0.00576529144041711
				0.0351843720888320 0.0377863982500000 0.00603412714032173
				0.0281474976710656 0.0307636524200000 0.00488830774578726
				0.0225179981368525 0.0227620396200000 0.00333842184361413
				0.0180143985094820 0.0168419733700000 0.00268276345751074
				0.0144115188075856 0.0139285108200000 0.00222217904193812
				0.0115292150460685 0.0118605612700000 0.00178005332214487
				};
			\addplot[only marks, mark=x, mark size=2pt, color of colormap={300}]
				plot[error bars/.cd, y dir=both, y explicit]
				table[x=eps, y=error, y error=error_delta]
				{
				eps error error_delta
				1 1.40372669494046 0.0824145081142738
				0.800000000000000 1.11995250004290 0.0884705300583424
				0.640000000000000 0.893898774499000 0.0760194454643851
				0.512000000000000 0.787839983923000 0.0665196125136270
				0.409600000000000 0.531139498826000 0.0530989833224664
				0.327680000000000 0.377207480110000 0.0462407877249660
				0.262144000000000 0.226433735130000 0.0351182078209684
				0.209715200000000 0.209581089330000 0.0287195901336503
				0.167772160000000 0.151631804420000 0.0235681083473183
				0.134217728000000 0.121504748920000 0.0173537415216022
				0.107374182400000 0.0954274174700000 0.0161534176288046
				0.0858993459200001 0.0882687074599999 0.0131027887668179
				0.0687194767360000 0.0566083970300000 0.00827897757267499
				0.0549755813888000 0.0473961147400000 0.00714402135098646
				0.0439804651110400 0.0348170940200000 0.00551421869465726
				0.0351843720888320 0.0331904162700000 0.00442605476659735
				0.0281474976710656 0.0242285586600000 0.00345739891504572
				0.0225179981368525 0.0214802208900000 0.00361059204853812
				0.0180143985094820 0.0150197971700000 0.00221631895075742
				0.0144115188075856 0.0124065590900000 0.00175119501951064
				0.0115292150460685 0.00982867712000001 0.00137505897496311
				0.00922337203685479 0.00793969447000000 0.00120355113893455
				0.00737869762948383 0.00653041485000000 0.000994504926264836
				0.00590295810358706 0.00465842422000002 0.000726180395860647
				0.00472236648286965 0.00491334197000000 0.000696939298454315
				};
			\addplot[only marks, mark=x, mark size=2pt, color of colormap={600}]
				plot[error bars/.cd, y dir=both, y explicit]
				table[x=eps, y=error, y error=error_delta]
				{
				eps error error_delta
				1 1.37597728509700 0.0978087745156142
				0.800000000000000 1.05945096290400 0.0984786143200493
				0.640000000000000 0.886658353740000 0.0866121889492767
				0.512000000000000 0.818213464900000 0.0744349160992445
				0.409600000000000 0.643496610540000 0.0672454105549025
				0.327680000000000 0.512428691900000 0.0638901859695035
				0.262144000000000 0.385556558940000 0.0505927922671960
				0.209715200000000 0.302991937160000 0.0350852538990343
				0.167772160000000 0.220378624550000 0.0301106708229267
				0.134217728000000 0.174059036980000 0.0258536129325492
				0.107374182400000 0.137750808600000 0.0179729170537635
				0.0858993459200001 0.0914479284600000 0.0139812416117793
				0.0687194767360000 0.0799986713400000 0.0111713919439875
				0.0549755813888000 0.0607137489800000 0.00870656183270174
				0.0439804651110400 0.0546123703900000 0.00774679469095545
				0.0351843720888320 0.0383273383900000 0.00606221717132340
				0.0281474976710656 0.0309869770200000 0.00485258380045250
				0.0225179981368525 0.0241924708700000 0.00363744413950662
				0.0180143985094820 0.0203458978600000 0.00296899123596728
				0.0144115188075856 0.0151127835300000 0.00251018122976063
				0.0115292150460685 0.0112933818800000 0.00168148974495163
				0.00922337203685479 0.00984538217999999 0.00141924015498561
				0.00737869762948383 0.00690450247000001 0.00102085643213254
				0.00590295810358706 0.00613576063999999 0.000887530552108717
				0.00472236648286965 0.00434193445000001 0.000693481226288422
				0.00377789318629572 0.00367583326000000 0.000519337081985985
				0.00302231454903658 0.00329234995000000 0.000443661665430249
				0.00241785163922926 0.00260172126000001 0.000364621118015930
				0.00193428131138341 0.00196506195000001 0.000296432616530746
				0.00154742504910673 0.00166866582999997 0.000232556847747214
				0.00123794003928538 0.00137025276000000 0.000211040649235300
				0.000990352031428306 0.00104601317999998 0.000178751091157049
				0.000792281625142645 0.000797655740000005 0.000114513132712890
				};
			\addplot[only marks, mark=x, mark size=2pt, color of colormap={875}]
				plot[error bars/.cd, y dir=both, y explicit]
				table[x=eps, y=error, y error=error_delta]
				{
				eps error error_delta
				1 0.927750547495000 0.0706255885984242
				0.800000000000000 0.718029276145000 0.0659266588222826
				0.640000000000000 0.594891622892000 0.0624770006018154
				0.512000000000000 0.537515523856000 0.0544568372481612
				0.409600000000000 0.405010687920000 0.0486580518949063
				0.327680000000000 0.345066154380000 0.0385272551594060
				0.262144000000000 0.253344899710000 0.0333530957992743
				0.209715200000000 0.300699554050000 0.0324996232997495
				0.167772160000000 0.194613242050000 0.0254600637794430
				0.134217728000000 0.193590862920000 0.0225836682736859
				0.107374182400000 0.120448967780000 0.0162082521559416
				0.0858993459200001 0.102729195730000 0.0121401879877038
				0.0687194767360000 0.0633219758400000 0.00950612189029575
				0.0549755813888000 0.0480192592200000 0.00691644948709024
				0.0439804651110400 0.0385373405400000 0.00543062328810570
				0.0351843720888320 0.0355859958000000 0.00533593225010195
				0.0281474976710656 0.0296901301900000 0.00407471697293018
				0.0225179981368525 0.0191786398800000 0.00265555798448199
				0.0180143985094820 0.0172024999400000 0.00207082833625228
				0.0144115188075856 0.0121603975000000 0.00180029388364868
				0.0115292150460685 0.0131702541200000 0.00184515587622973
				0.00922337203685479 0.00871229401000000 0.00143618082418338
				0.00737869762948383 0.00849305160000002 0.00109177040635769
				0.00590295810358706 0.00564543769000000 0.000697827019610954
				0.00472236648286965 0.00501318780999999 0.000656435611371164
				0.00377789318629572 0.00330409512000000 0.000527599842245410
				0.00302231454903658 0.00281851899000001 0.000415694591651150
				0.00241785163922926 0.00205091878000001 0.000327316915456641
				0.00193428131138341 0.00192462175000001 0.000297094172867742
				0.00154742504910673 0.00139322307000000 0.000249076109556646
				0.00123794003928538 0.00120632333000001 0.000167333038485815
				0.000990352031428306 0.000892756600000013 0.000136763245043851
				0.000792281625142645 0.000613424469999993 9.49177647190577e-05
				0.000633825300114116 0.000591506520000025 8.31355426240431e-05
				};
			\addplot[no markers, dashed, black, domain=0.4*10^(-3):10^(-1)]{x};
			\legend{$\gamma=0.45$ (eq),$\gamma=0.25$ (eq),$\gamma=0.45$ (ad),$\gamma=0.25$ (ad)}
		\end{loglogaxis}
	\end{tikzpicture}
	\caption{Error given by \eqref{eq:error-quant} vs.\ error tolerance $\varepsilon$
		for the quantile approximation of $S(w|B|)$ with $q=0.95$.}
	\label{fig:quant:2}
\end{figure}

In Figure \ref{fig:quant:2} we relate the actual error
to the error tolerance $\varepsilon$. We see that
both algorithms almost perfectly achieve the computational goal, namely,
an error close to the input $\varepsilon$.

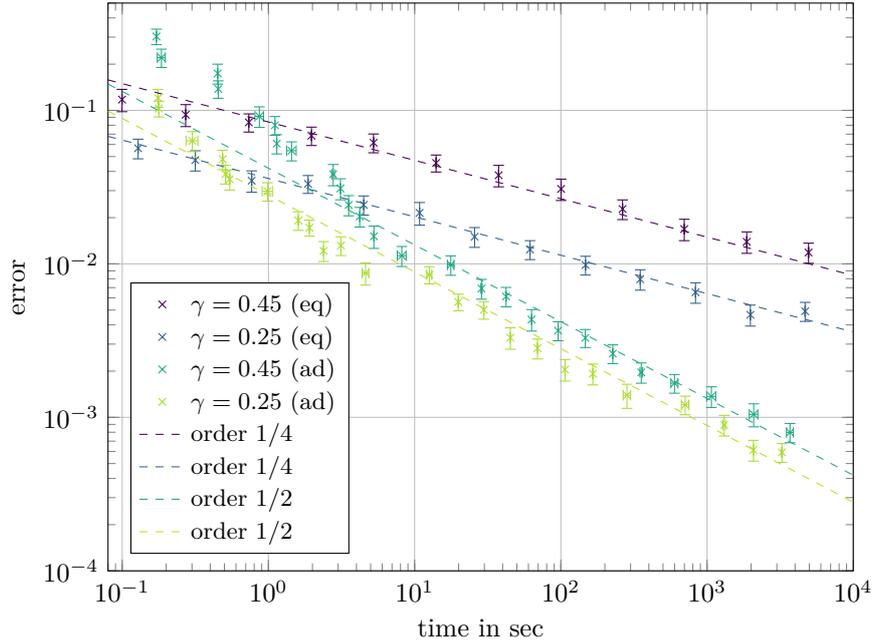
\begin{figure}[htb]
	\centering
	\begin{tikzpicture}
		\begin{loglogaxis}[
				xmin = 0.8*10^(-1), xmax = 10^4, ymin = 10^(-4), ymax = 0.5*10^(0),
				xlabel = {time in sec}, ylabel = {error},
				legend style = {font=\small}, legend pos = south west, legend cell align = left,
				width = 0.9\textwidth,
				height = 0.72\textwidth,
				grid = major,
				colormap name = viridis
			]
			\addplot[only marks, mark=x, mark size=2pt, color of colormap={0}]
				plot[error bars/.cd, x dir=both, x explicit, y dir=both, y explicit]
				table[x=time, y=error, x error=time_delta, y error=error_delta]
				{
				time time_delta error error_delta
				0.0996774197400000 7.50150093863392e-05 0.117503663360000 0.0194940202683091
				0.271736421680000 0.000598395749411140 0.0936704797600000 0.0152909661297684
				0.734496944720000 0.000787189996554897 0.0835105300600000 0.0113085949088017
				1.97073449755000 0.00130077153383195 0.0683912413300000 0.00938525443405078
				5.23571686810000 0.00238860241356142 0.0615366029000000 0.00863366371607376
				14.0235998280000 0.00565751583038700 0.0454179828000000 0.00576529144041711
				37.5568777404000 0.00911541481987509 0.0377863982500000 0.00603412714032173
				100.404032490000 0.0152764703479056 0.0307636524200000 0.00488830774578726
				264.279521180000 0.0261795205306726 0.0227620396200000 0.00333842184361413
				698.412779048000 0.0556903850073412 0.0168419733700000 0.00268276345751074
				1868.94733780000 5.17474255061454 0.0139285108200000 0.00222217904193812
				4943.47330406000 26.4963519849608 0.0118605612700000 0.00178005332214487
				};
			\addplot[only marks, mark=x, mark size=2pt, color of colormap={300}]
				plot[error bars/.cd, x dir=both, x explicit, y dir=both, y explicit]
				table[x=time, y=error, x error=time_delta, y error=error_delta]
				{
				time time_delta error error_delta
				0.128422622260000 0.000201620075245839 0.0566083970300000 0.00827897757267499
				0.316162201600000 0.000600781577090590 0.0473961147400000 0.00714402135098646
				0.768207514190000 0.000837310176237373 0.0348170940200000 0.00551421869465726
				1.86660326007000 0.00117944827859564 0.0331904162700000 0.00442605476659735
				4.48546949404000 0.00175393532447760 0.0242285586600000 0.00345739891504572
				10.8101637020000 0.00499711512221215 0.0214802208900000 0.00361059204853812
				25.7223616709000 0.00781710156478237 0.0150197971700000 0.00221631895075742
				61.5496125205000 0.00933804456871292 0.0124065590900000 0.00175119501951064
				147.495240863000 0.0162121306878344 0.00982867712000001 0.00137505897496311
				349.128735853000 0.0313581073114384 0.00793969447000000 0.00120355113893455
				834.131722742000 0.0626081263407587 0.00653041485000000 0.000994504926264836
				1978.88377252000 4.17993679488181 0.00465842422000002 0.000726180395860647
				4685.42464921000 26.4893402089611 0.00491334197000000 0.000696939298454315
				};
			\addplot[only marks, mark=x, mark size=2pt, color of colormap={600}]
				plot[error bars/.cd, x dir=both, x explicit, y dir=both, y explicit]
				table[x=time, y=error, x error=time_delta, y error=error_delta]
				{
				time time_delta error error_delta
				0.171404205270000 0.000235221389360508 0.302991937160000 0.0350852538990343
				0.185506245120000 0.0118497559612979 0.220378624550000 0.0301106708229267
				0.449846659540000 0.000105481512294341 0.174059036980000 0.0258536129325492
				0.453591039280000 0.000141749013948592 0.137750808600000 0.0179729170537635
				0.867342376240000 0.0589538653856125 0.0914479284600000 0.0139812416117793
				1.10753778685000 0.000206072011954216 0.0799986713400000 0.0111713919439875
				1.14056889265000 0.000662299217955102 0.0607137489800000 0.00870656183270174
				1.44039179990000 0.108545663743354 0.0546123703900000 0.00774679469095545
				2.77280667580000 0.0604436134272873 0.0383273383900000 0.00606221717132340
				3.11039693724000 0.000614763420811722 0.0309869770200000 0.00485258380045250
				3.54139664856000 0.000677079694682755 0.0241924708700000 0.00363744413950662
				4.21382889118000 0.000752892084692891 0.0203458978600000 0.00296899123596728
				5.26564793050000 0.000774629825082082 0.0151127835300000 0.00251018122976063
				8.17280451192000 0.570293862977537 0.0112933818800000 0.00168148974495163
				17.6740110812600 0.907718414189757 0.00984538217999999 0.00141924015498561
				28.6755021498000 0.300976739609814 0.00690450247000001 0.00102085643213254
				42.2036969286000 0.00433880940587072 0.00613576063999999 0.000887530552108717
				63.1152804071000 0.00540600921498137 0.00434193445000001 0.000693481226288422
				95.7710005333000 0.00585213848703352 0.00367583326000000 0.000519337081985985
				146.890329493000 0.0118568552063574 0.00329234995000000 0.000443661665430249
				226.604220431000 0.00794230477127005 0.00260172126000001 0.000364621118015930
				354.953871935000 7.43247197658241 0.00196506195000001 0.000296432616530746
				598.919634606000 33.2298517296405 0.00166866582999997 0.000232556847747214
				1070.81024480900 77.3569643757978 0.00137025276000000 0.000211040649235300
				2084.43683398000 140.738203991551 0.00104601317999998 0.000178751091157049
				3691.46999325000 201.011737860943 0.000797655740000005 0.000114513132712890
				};
			\addplot[only marks, mark=x, mark size=2pt, color of colormap={875}]
				plot[error bars/.cd, x dir=both, x explicit, y dir=both, y explicit]
				table[x=time, y=error, x error=time_delta, y error=error_delta]
				{
				time time_delta error error_delta
				0.175898682020000 0.000277624135349938 0.120448967780000 0.0162082521559416
				0.178097902430000 9.80306453072588e-05 0.102729195730000 0.0121401879877038
				0.301407743530000 0.0281370554418174 0.0633219758400000 0.00950612189029575
				0.486311506110000 0.000415918510840946 0.0480192592200000 0.00691644948709024
				0.509069775700000 0.000446058228036292 0.0385373405400000 0.00543062328810570
				0.543660371070000 0.000435596135410306 0.0355859958000000 0.00533593225010195
				0.985993681290000 0.0795338135103093 0.0296901301900000 0.00407471697293018
				1.60150750676000 0.000999183966301126 0.0191786398800000 0.00265555798448199
				1.90635982657000 0.000844478515039197 0.0172024999400000 0.00207082833625228
				2.38535830084000 0.00180226688337316 0.0121603975000000 0.00180029388364868
				3.13260960775000 0.00163943951473442 0.0131702541200000 0.00184515587622973
				4.61697288052000 0.245642443523683 0.00871229401000000 0.00143618082418338
				12.6037323503400 0.493655379515006 0.00849305160000002 0.00109177040635769
				19.8949880128000 0.00737526872756871 0.00564543769000000 0.000697827019610954
				29.7417980759000 0.00712875633771894 0.00501318780999999 0.000656435611371164
				45.1594949159000 0.00874349927362431 0.00330409512000000 0.000527599842245410
				69.1893009924000 0.00906583376642261 0.00281851899000001 0.000415694591651150
				106.823543168000 0.0137110464633404 0.00205091878000001 0.000327316915456641
				165.501395294000 0.0215791175563706 0.00192462175000001 0.000297094172867742
				283.420999801000 16.3831096683491 0.00139322307000000 0.000249076109556646
				708.764705913000 41.6178928934785 0.00120632333000001 0.000167333038485815
				1311.12388538300 27.6887224566396 0.000892756600000013 0.000136763245043851
				2076.37464660000 0.114856376385892 0.000613424469999993 9.49177647190577e-05
				3253.79045818000 26.4930408188466 0.000591506520000025 8.31355426240431e-05
				};
			\addplot[no markers, dashed, color of colormap={0},   domain=0.8*10^(-1):10^4]{0.084*x^(-0.25)};
			\addplot[no markers, dashed, color of colormap={300}, domain=0.8*10^(-1):10^4]{0.036*x^(-0.25)};
			\addplot[no markers, dashed, color of colormap={600}, domain=0.8*10^(-1):10^4]{0.042*x^(-0.5)};
			\addplot[no markers, dashed, color of colormap={900}, domain=0.8*10^(-1):10^4]{0.028*x^(-0.5)};
			\legend{$\gamma=0.45$ (eq),$\gamma=0.25$ (eq),$\gamma=0.45$ (ad),$\gamma=0.25$ (ad),
				order $1/4$, order $1/4$, order $1/2$, order $1/2$}
		\end{loglogaxis}
	\end{tikzpicture}
	\caption{Error given by \eqref{eq:error-quant} vs.\ average run-time
		for the quantile approximation of $S(w|B|)$ with $q=0.95$.}
	\label{fig:quant:1}
\end{figure}

As for the strong approximation, the relation between the error
and the average run-time is most important, see
Figure \ref{fig:quant:1} for the results.
For $Q^\adap_{\varepsilon}(w, q)$ we observe a polynomial order of
convergence of about $1/2$, while the corresponding order for
$Q^\equi_{\varepsilon}(w, q)$ is only about $1/4$. This corresponds
to the findings from Section \ref{s3} and to Remark \ref{r7}:
The orders of convergence of $A_n^\adap(w,\cdot)$ and
$A^\equi_n(w,\cdot)$ for the strong approximation of $S(w|B|)$
are given by $\infty$ and $1/2$, respectively,
see Figure~\ref{fig:strong:3}, and the order of convergence for the
quantile approximation should be $1/2$, see Remark~\ref{r7}.

The algorithm $Q^\adap_{\varepsilon}(w, q)$ achieves an error $10^{-2}$
in $5$ seconds for $\gamma=0.25$ and in $12$ seconds for $\gamma =
0.45$, and even an error $10^{-3}$ in less than $15$ or $35$
minutes, respectively.
The algorithm $Q^\equi_{\varepsilon}(w, q)$ achieves the error $10^{-2}$
in $2$ minutes for $\gamma=0.25$ and
(based on a extrapolation beyond the range of our numerical experiment)
in more than $2$ hours for $\gamma = 0.45$; the corresponding
run-times for the error $10^{-3}$ are
6 days and 6 years, respectively.

\subsection*{Acknowledgement}

The authors are grateful to
Fabian Schlechter for programming and computer support.
The authors thank
Jan Henrik Fitschen,
Yannick Kreis,
and
Lukas Mayer
for valuable discussions.
Stefanie Schwaar was supported by the Deutsche Forschungsgemeinschaft
(DFG) within the RTG 1932
`Stochastic Models for Innovations in the Engineering Sciences'.

\bibliography{AdaptiveQuantileComputation}

\begin{thebibliography}{31}
\providecommand{\natexlab}[1]{#1}
\providecommand{\url}[1]{\texttt{#1}}
\expandafter\ifx\csname urlstyle\endcsname\relax
  \providecommand{\doi}[1]{doi: #1}\else
  \providecommand{\doi}{doi: \begingroup \urlstyle{rm}\Url}\fi

\bibitem[Akashi et~al.(2018)Akashi, Dette, and Liu]{AkashiEtAl18}
F.~Akashi, H.~Dette, and Y.~Liu.
\newblock Change-point detection in autoregressive models with no moment
  assumptions.
\newblock \emph{Journal of Time Series Analysis}, 39\penalty0 (5):\penalty0
  763--786, 2018.

\bibitem[Asmussen et~al.(1995)Asmussen, Glynn, and Pitman]{MR1384357}
S.~Asmussen, P.~Glynn, and J.~Pitman.
\newblock Discretization error in simulation of one-dimensional reflecting
  {B}rownian motion.
\newblock \emph{Ann.\ Appl.\ Probab.}, 5\penalty0 (4):\penalty0 875--896, 1995.

\bibitem[Aue and Horv\'{a}th(2013)]{AueHorvath13}
A.~Aue and L.~Horv\'{a}th.
\newblock Structural breaks in time series.
\newblock \emph{J. Time Series Anal.}, 34\penalty0 (1):\penalty0 1--16, 2013.

\bibitem[Aue et~al.(2018)Aue, Rice, and S\"onmez]{Aue_et_al_2018}
A.~Aue, G.~Rice, and O.~S\"onmez.
\newblock Detecting and dating structural breaks in functional data without
  dimension reduction.
\newblock \emph{J. R. Statist.\ Soc.\ B}, 80\penalty0 (3):\penalty0 509--529,
  2018.

\bibitem[Bickel and Rosenblatt(1973)]{BickRos73}
P.~Bickel and M.~Rosenblatt.
\newblock On some global measures of the deviations of density function
  estimates.
\newblock \emph{Ann.\ Statist.}, 1\penalty0 (3):\penalty0 1071--1091, 1973.

\bibitem[Borodin and Salminen(2002)]{borodin_salminen_1996}
A.~Borodin and P.~Salminen.
\newblock \emph{Handbook of {B}rownian Motion}.
\newblock Springer, $2^{\text{nd}}$ edition, 2002.

\bibitem[Calvin(1997)]{Calvin1997}
J.~M. Calvin.
\newblock Average performance of a class of adaptive algorithms for global
  optimization.
\newblock \emph{Ann.\ Appl.\ Probab.}, 7:\penalty0 711--730, 1997.

\bibitem[Calvin(2001)]{Calvin2001}
J.~M. Calvin.
\newblock A one-dimensional optimization algorithm and its convergence rate
  under the {W}iener measure.
\newblock \emph{J. Complexity}, 17\penalty0 (2):\penalty0 306--344, 2001.

\bibitem[Calvin(2004)]{Calvin2004}
J.~M. Calvin.
\newblock Lower bound on complexity of optimization of continuous functions.
\newblock \emph{J. Complexity}, 20\penalty0 (5):\penalty0 773--795, 2004.

\bibitem[Calvin et~al.(2017)Calvin, Hefter, and Herzwurm]{MR3605752}
J.~M. Calvin, M.~Hefter, and A.~Herzwurm.
\newblock Adaptive approximation of the minimum of {B}rownian motion.
\newblock \emph{J. Complexity}, 39:\penalty0 17--37, 2017.

\bibitem[Cormen et~al.(2003)Cormen, Leiserson, Rivest, and
  Stein]{cormen2009introduction}
T.~H. Cormen, C.~E. Leiserson, R.~L. Rivest, and C.~Stein.
\newblock \emph{Introduction to Algorithms}.
\newblock MIT Press, $4^{\text{th}}$ edition, 2003.

\bibitem[Cs\"org\H{o} and Horv\'ath(1988)]{CsorgoHorvath88}
M.~Cs\"org\H{o} and L.~Horv\'ath.
\newblock Nonparametric methods for changepoint problems.
\newblock In \emph{Quality Control and Reliability}, volume~7 of \emph{Handbook
  of Statistics}, pages 403--425. Elsevier, 1988.

\bibitem[Cs{\"o}rg\H{o} and Horv{\'a}th(1997)]{book:CsorgoHorvath}
M.~Cs{\"o}rg\H{o} and L.~Horv{\'a}th.
\newblock \emph{Limit Theorems in Change-Point Analysis}.
\newblock Wiley series in probability and statistics. John Wiley \& Sons Ltd.,
  1997.

\bibitem[David and Nagaraja(2003)]{MR1994955}
H.~A. David and H.~N. Nagaraja.
\newblock \emph{Order Statistics}.
\newblock Wiley Series in Probability and Statistics. Wiley-Interscience,
  Hoboken, NJ, third edition, 2003.

\bibitem[DeLong(1981)]{DeLong81}
D.~M. DeLong.
\newblock Crossing probabilities for a square root boundary by a {B}essel
  process.
\newblock \emph{Commun.\ Statist.-Theor.\ Meth.}, 10\penalty0 (21):\penalty0
  2197--2213, 1981.

\bibitem[Eastwood and Eastwood(1998)]{EastwoodEastwood98}
B.~J. Eastwood and V.~R. Eastwood.
\newblock Tabulating weighted sup-norm functionals of {B}rownian bridges via
  {M}onte {C}arlo simulation.
\newblock In \emph{Asymptotic Methods in Probability and Statistics}, pages
  707--719. North-Holland, 1998.

\bibitem[Fried and Imhoff(2004)]{FriedImhoff04}
R.~Fried and M.~Imhoff.
\newblock On the online detection of monotonic trends in time series.
\newblock \emph{Biometrical Journal}, 46\penalty0 (1):\penalty0 90--102, 2004.

\bibitem[Gallagher et~al.(2013)Gallagher, Lund, and Robbins]{GallagherEtAl13}
C.~Gallagher, R.~Lund, and M.~Robbins.
\newblock Changepoint detection in climate time series with long-term trends.
\newblock \emph{Journal of Climate}, 26\penalty0 (14):\penalty0 4994--5006,
  2013.

\bibitem[H{\"a}rdle(1990)]{book:Haerdle90}
W.~H{\"a}rdle.
\newblock \emph{Applied Nonparametric Regression}.
\newblock Cambridge University Press, Cambridge, 1990.

\bibitem[Hefter and Herzwurm(2017)]{MR3744680}
M.~Hefter and A.~Herzwurm.
\newblock Optimal strong approximation of the one-dimensional squared {B}essel
  process.
\newblock \emph{Commun.\ Math.\ Sci.}, 15\penalty0 (8):\penalty0 2121--2141,
  2017.

\bibitem[Hochberg(1988)]{Hochberg88}
Y.~Hochberg.
\newblock A sharper {B}onferroni procedure for multiple tests of significance.
\newblock \emph{Biometrika}, 80\penalty0 (4):\penalty0 800--802, 1988.

\bibitem[Horváth and Kokoszka(2012)]{book:HorvKoko12}
L.~Horváth and P.~Kokoszka.
\newblock \emph{Inference for Functional Data with Applications}.
\newblock Springer, Berlin-Heidelberg-New York, 2012.

\bibitem[Kirch(2006)]{dissertationKirch}
C.~Kirch.
\newblock \emph{Resampling Methods for the Change Analysis of Dependent Data}.
\newblock PhD thesis, Universit{\"a}t zu K{\"o}ln, Juni 2006.

\bibitem[Kirch and Tadjuidje~Kamgaing(2016)]{incoll:KirchTadjuidje16}
C.~Kirch and J.~Tadjuidje~Kamgaing.
\newblock Detection of change points in discrete-valued time series.
\newblock In \emph{Handbook of Discrete-valued Time Series}, pages 219--244.
  CRC Press, 2016.

\bibitem[Orasch and Pouliot(2004)]{OraschPouliot04}
M.~Orasch and W.~Pouliot.
\newblock Tabulating weighted sup-norm functionals used in change-point
  analysis.
\newblock \emph{J. Stat. Comput. Simul.}, 74\penalty0 (4):\penalty0 249--276,
  2004.

\bibitem[Page(1957)]{Page57}
E.~S. Page.
\newblock On problems in which a change in a parameter occurs at an unknown
  point.
\newblock \emph{Biometrika}, 44\penalty0 (1/2):\penalty0 248--252, 1957.

\bibitem[Ritter(1990)]{MR1085383}
K.~Ritter.
\newblock Approximation and optimization on the {W}iener space.
\newblock \emph{J. Complexity}, 6\penalty0 (4):\penalty0 337--364, 1990.

\bibitem[Salminen and Yor(2011)]{salminen_yor_2011}
P.~Salminen and M.~Yor.
\newblock On hitting times of affine boundaries by reflecting {B}rownian motion
  and {B}essel processes.
\newblock \emph{Period.\ Math.\ Hungar.}, 62\penalty0 (1):\penalty0 75--101,
  2011.

\bibitem[Schwaar(2016)]{Dis:Schwaar}
S.~Schwaar.
\newblock \emph{Asymptotics for change-point tests and change-point
  estimators}.
\newblock PhD thesis, Technische Universit\"{a}t Kaiserslautern, 2016.

\bibitem[Schwaar(2020)]{Schwaar20}
S.~Schwaar.
\newblock A data-driven change-point estimator.
\newblock \emph{arXiv:2010.12449}, 2020.

\bibitem[Yor(1984)]{MR777516}
M.~Yor.
\newblock On square-root boundaries for {B}essel processes, and pole-seeking
  {B}rownian motion.
\newblock In \emph{Stochastic {A}nalysis and {A}pplications}, volume 1095 of
  \emph{Lecture Notes in Math.}, pages 100--107. Springer, 1984.

\end{thebibliography}
\bibliographystyle{abbrvnat}

\end{document}